\documentclass[pra,twocolumn,showpacs]{revtex4}
\usepackage{graphicx}
\usepackage{epsfig}
\usepackage[usenames]{color}

\newcommand{\beq}{\begin{equation}}
\newcommand{\eeq}{\end{equation}}
\newcommand{\beqa}{\begin{eqnarray}}
\newcommand{\eeqa}{\end{eqnarray}}
\newcommand{\ba}{\begin{array}}
\newcommand{\ea}{\end{array}}

\begin{document}

\title{Extended Thomas-Fermi Density Functional \\ 
for the Unitary Fermi Gas} 
\author{Luca Salasnich and Flavio Toigo}
\affiliation{Department of Physics ``Galileo Galilei'', 
CNISM and CNR-INFM, 
University of Padua, Via Marzolo 8, 35131 Padua, Italy} 

\begin{abstract}
We determine the energy density $\xi (3/5) n \varepsilon_F$ 
and the gradient correction $\lambda \hbar^2(\nabla n)^2/(8m\, n)$ 
of the extended Thomas-Fermi (ETF) density functional, where $n$ is 
number density and $\varepsilon_F$ is Fermi energy,  
for a trapped two-components 
Fermi gas with infinite scattering length (unitary Fermi gas)
on the basis of recent diffusion Monte Carlo (DMC) 
calculations [Phys. Rev. Lett. {\bf 99}, 233201 (2007)]. 
In particular we find that $\xi=0.455$ and $\lambda=0.13$ give 
the best fit of the DMC data with an even number $N$ of particles. 
We also study the odd-even splitting 
$\gamma N^{1/9} \hbar \omega$ of the ground-state energy 
for the unitary gas in a harmonic trap of frequency $\omega$
determining the constant $\gamma$. Finally we investigate the effect 
of the gradient term in the time-dependent ETF model 
by introducing generalized Galilei-invariant hydrodynamics equations. 
\end{abstract}


\maketitle

\section{Introduction} 

The crossover from the weakly paired Bardeen-Cooper-Schrieffer 
(BCS) state to the Bose-Einstein condensate (BEC) 
of molecular dimers with ultra-cold two-hyperfine-components Fermi
vapors atoms has been investigated 
by several experimental groups with $^{40}$K atoms 
\cite{greiner,regal,kinast} 
and $^6$Li atoms \cite{zwierlein,chin}. 
In the unitary limit of infinite scattering length \cite{uexp} 
obtained by tuning an external background magnetic field 
near a Feshbach resonance \cite{stringa-fermi}, 
the Fermi superfluid exhibits universal properties 
\cite{stringa-fermi,baker,mc1,mc2}. 

It has been suggested that at zero-temperature 
the unitary Fermi gas 
can be described by the density functional theory (DFT) 
\cite{bulgac,kim,manini05,papenbrock,zyl,son,sala-josephson,rupak}. 
Bulgac and Yu \cite{bulgac} introduced a superfluid DFT (SDFT) based 
on a Bogoliubov-de Gennes (BdG) approach to superfluid fermions, 
in the same spirit as the density functional formulation for 
superconductors \cite{oliveira}. 
Papenbrock and Bhattacharyya \cite{papenbrock} have instead proposed 
a Kohn-Sham (KS) density functional with an effective mass 
to take into account nonlocality effects. 
To treat nonuniform systems, other authors 
\cite{kim,manini05,son,sala-josephson,rupak} have added 
a gradient term to the leading Thomas-Fermi energy, since such a term
is surely necessary when surfaces are present 
\cite{von,kirzhnits}, at least in three spatial dimensions \cite{nota}. 
An energy functional
for fermions not written in terms of single-particle orbitals, but 
only in terms of the density and its derivatives 
is usually called extended Thomas-Fermi (ETF) functional 
\cite{lipparini,ring}. 
It may be seen as an effective field theory where
the gradient correction $\lambda \hbar^2(\nabla n)^2/(8m\, n)$ 
can be interpreted as the next-to-leading 
term \cite{son,rupak}, with $n({\bf r})$ the local number density 
and $m$ the atomic mass. 

We wish to point out that both the energy functionals 
proposed by Bulgac and Yu \cite{bulgac} 
and Papenbrock and Bhattacharyya \cite{papenbrock} are functionals of 
the density through single particle orbitals
(the BdG or KS orbitals). Therefore they can be used in actual numerical 
calculations 
only when the number of fermions is small, since 
they require a self consistent calculation of single-particle states 
whose number increases linearly with the number of particles.

On the contrary, one encounters no limitation in the number of particles 
which may be treated through ETF functionals, 
since in this case the functional 
depends only on a single function of the coordinate, i.e. the particle density.
Of course one trades simplicity with accuracy: while the BdG and KS schemes 
are built to account for the main contribution to the kinetic energy, and 
treat it exactly in noninteracting systems even with a non-uniform density 
varying in space, the TF approach gives the exact kinetic energy only for a 
uniform system and even when extended with the addition of gradient and 
higher order derivatives of the density, the ETF functional is not able to 
reproduce shell effects in the density profile \cite{lipparini,ring}. 

In spite of this limitation, but in the light of the great simplification
introduced in numerical calculations, we believe that it is useful to 
analyze the ETF approximation and comment on its dynamical generalization, 
which amounts to introducing a quantum pressure term 
$-\lambda {\hbar^2}\nabla^2 \sqrt{n}/(2m\sqrt{n})$ 
into the hydrodynamic equations of superfluids.

The value of the coefficient $\lambda$ is debated. In the papers of 
Kim and Zubarev \cite{kim} 
and Manini and Salasnich \cite{manini05} the authors set $\lambda=1$ 
over the full BCS-BEC crossover. More recently we have suggested 
$\lambda=1/4$ \cite{sala-josephson,sala-new}. 
This suggestion is in good agreement 
with a theoretical estimate 
based on an epsilon expansion around $d=4-\varepsilon$ 
spatial dimensions in the unitary regime \cite{rupak}. 

In this paper we comment on the ETF for a two-components Fermi gas at 
unitarity and determine its  parameters by fitting recent Monte Carlo results
\cite{chang,blume} for
the energy  of fermions confined in a spherical harmonic trap of frequency 
$\omega$ in this regime. 

Since the interaction potential does not introduce any new length, 
the universal contribution to the energy density in an ETF functional 
appropriate to a spin balanced Fermi liquid at unitarity may  be considered, 
in its simplest form, as the sum of a term proportional ($\xi$ being the 
constant of proportionality) to the energy density of a uniform noninteracting 
system with the same density and  of a term containing the gradient of the 
density with a coefficient 
$\lambda$ meant to take into account phenomenologically also higher order 
derivatives \cite{parr}.

By minimizing such ETF for a fixed number of particles, we find that 
the values  $\xi=0.455$ and $\lambda=0.13$ give the closest results to 
the Monte Carlo energies of a fully superfluid system with an even 
number $N$ of fermions confined in a harmonic well at unitarity, 
as calculated by Ref. \cite{blume}.
When treating systems with an odd number of particles, we must correct 
the calculated value of the ETF ground state energies corresponding to 
these parameters to account for the presence of the unpaired particle. 
According to Son \cite{son2}, for fermions confined by a harmonic potential, 
the correction depends on the number of particle and takes  the form
$\Delta E = \gamma N^{1/9}$ (in units of $ \hbar \omega$). We then find that 
 $\gamma = 0.856$ provides the best fit to the DMC data \cite{blume}.  
In section IV,  we investigate the effect of the gradient 
term on the dynamics of the Fermi superfluid by introducing 
generalized hydrodynamics equations and a Galilei-invariant nonlinear 
Schr\"odinger equation of the Guerra-Pusterla type \cite{guerra,doebner}
which is fully equivalent to them. 

\section{Extended Thomas-Fermi functional} 

Let us consider an interacting Fermi gas trapped 
by a potential $U({\bf r})$. 
Its TF energy functional  is:
\beq 
E_{TF} = \int  d^3{\bf r} \ 
n({\bf r}) \Big[  {\varepsilon}(n({\bf r})) 
+ U({\bf r}) \Big]  \ , 
\label{e-lda}
\eeq 
where ${\varepsilon}(n)$ is the energy per particle
of a uniform Fermi system with density $n$ equal to the local density 
$n({\bf r})$
of fermions. The total number of fermions is 
\beq
N = \int d^3{\bf r} \ n({\bf r})   \; . 
\label{norma}
\eeq
By minimizing $E_{TF}$ with respect to the density $n({\bf r})$,  with 
the constraint of a fixed number of particles,  one finds 
\beq
\mu(n({\bf r})) + U({\bf r}) = \bar{\mu} \; , 
\label{chem-lda}
\eeq 
where $\mu(n)={\partial (n {\varepsilon}(n))\over \partial n}$ is the 
bulk chemical potential of a uniform system of density $n$ and $\bar{\mu}$ 
is the chemical 
potential of the non uniform system, i.e. the Lagrange 
multiplier fixed by the normalization (\ref{norma}). 

In the unitary limit, no characteristic length is set by the 
interatomic-potential since
its s-wave scattering length $a_F$  diverges ($a_F\to\pm\infty$).     
The energy per particle of a uniform two-spin components Fermi gas at 
unitarity must then depend only on $\hbar$, on the mass of fermions $m$, 
and on the only length characterizing the system, i.e. the average distance 
between particles $\propto n^{-1/3}$ \cite{stringa-fermi}.  It is usually 
written as:  
\beq 
{\varepsilon}(n) = \xi 
{3\over 5} {\hbar^2 \over 2m} (3\pi^2)^{2/3} n^{2/3} \; , 
\label{eos} 
\eeq 
where $\xi$ is a universal parameter which 
can be determined from ab-initio calculations.
Notice that:  $(\hbar^2/2m)(3\pi^2)^{2/3} n^{2/3}= \varepsilon_F$, where 
$\varepsilon_F$
is the Fermi energy of the ideal fermionic gas. Thus $\xi$ is simply the 
ratio between the energy per particle of the uniform interacting system 
at unitarity and the corresponding energy in a non-interacting system.
Monte Carlo calculations for a uniform unpolarized two-spin components Fermi 
gas suggest $\xi\simeq 0.45$; in particular, 
$\xi=0.42\pm 0.01$ according to \cite{mc1} and 
$\xi=0.44\pm 0.02$ according to \cite{mc2}. The bulk chemical 
potential associated to Eq. (\ref{eos}) is 
\beq 
\mu(n) = \xi {\hbar^2 \over 2m} (3\pi^2)^{2/3} n^{2/3} \; . 
\label{mu}
\eeq

If the system is confined by a spherically-symmetric 
harmonic potential :
\beq
U({\bf r})= {1\over 2} m \omega^2 r^2 \; ,
\label{harmonic}
\eeq
its density profile $n({\bf r})$  obtained  
from Eq. (\ref{chem-lda}) is :
\beq
n({\bf r}) = n(0) \left( 1 - {r^2\over r_F^2}\right)^{3/2} \; , 
\label{density-tf}
\eeq
where $n(0)=(2 m \bar{\mu})^{3/2}/(3\pi^2 \hbar^3 \xi^{3/2})$, 
$r_F=\sqrt{2\bar{\mu}/(m\omega^2)}$ and 
$\bar{\mu}=\hbar\omega \sqrt{\xi}(3N)^{1/3}$. 
Obviously, the expression (\ref{density-tf}) for the TF density profile of 
the unitary Fermi gas in a harmonic 
potential coincides with that of  an ideal Fermi gas 
\cite{sala-ideal}, but its parameters are modified by the presence of 
$\xi$ in the equation of state (\ref{eos}). 

As previously stressed the TF functional must be extended to take into 
account other characteristic lengths related to the spatial variations 
of the density, besides the average particle separation. As a consequence, 
the energy per particle must contain additional terms, which scale as the 
square of the inverse of  these various lengths. For this reason, as a 
simple approximation, we add to the energy per particle of Eq. (\ref{eos}), 
a term 
\beq 
\lambda {\hbar^2 \over 8 m} {\Big( {\nabla n\over n} \Big) }^2
= \lambda {\hbar^2\over 2m} {\Big( {\nabla \sqrt n \over \sqrt n} \Big) }^2 
\label{grad-term} \; , 
\eeq 
which may be seen as the  first term in a gradient expansion.
We notice that, according to the 
Kirzhnits expansion of the quantum kinetic operator 
in powers of $\hbar$ \cite{kirzhnits}, $\lambda$ must take the value 
$\lambda=1/9$ \cite{march,sala-gradient}
for an ideal, noninteracting, Fermi gas. 
Historically, a term of this form, but with $\lambda = 1$,  was introduced 
in a pioneering paper  \cite{von} by von Weizs\"acker  to treat surface 
effects in nuclei.

Instead than trying to determine it from first principles, we consider 
$\lambda$ as a phenomenological parameter accounting for 
the increase of kinetic energy due the entire spatial variation of the 
density.
We remark that this attitude is adopted in  many applications of Density 
Functional Theory 
to atomic and molecular physics where
the gradient term (\ref{grad-term}) takes into account phenomenologically, 
through $\lambda$, all the possible corrections of a 
gradient expansion \cite{parr}.  

The new energy functional reads \cite{kirzhnits,zaremba,march,sala-gradient} 
\beq 
E = \int d^3{\bf r} \ n({\bf r}) \Big[ {\varepsilon}_g(n({\bf r}),
\nabla n({\bf r})) 
+ U({\bf r})  \Big] \  \; ,  
\label{e-dft}
\eeq 
where 
\beq 
{\varepsilon}_g(n,\nabla n) = {\varepsilon}(n) + 
\lambda {\hbar^2 \over 8 m} {(\nabla n)^2\over n^2} 
\eeq 
is a generalized energy per particle  
which includes the $\hbar$-dependent gradient correction. 
 
We remark that in our ETF the constants $\xi$ and $\lambda$ are independent, 
implying that the ratios between the energies per particle corresponding 
to the two terms in  a unitary and in a nonininteracting fermi systems may 
be different. Equal ratios imply $\lambda = \xi /9$ \cite{kirzhnits}.
 
By minimizing the energy functional (\ref{e-dft}) with 
the constraint (\ref{norma}) one gets the partial 
differential equation obeyed by the ground state density:
\beq
\left[ \lambda {\hbar^2\over 2m} \nabla^2 
+ \mu(n({\bf r})) + U({\bf r}) \right] \sqrt{n({\bf r})} = 
\bar{\mu} \ \sqrt{n({\bf r})} \; .  
\label{chem-dft} 
\eeq 

For the study of hydrodynamics in Fermi superfluids Kim and Zubarev 
\cite{kim}, and also Manini and Salasnich \cite{manini05}, 
used $\lambda=1$ over the full BCS-BEC crossover. 
More recently we have suggested 
$\lambda=1/4$ \cite{sala-josephson} on the basis 
of the correct relationiship between phase and 
superfluid velocity \cite{sala-new}.  
This suggestion is in agreement
with the theoretical estimate of Rupak and Sch\"afer \cite{rupak}, 
obtained from an epsilon expansion 
around $d=4-\varepsilon$ spatial dimensions 
for a Fermi gas in the unitary regime. In particular, 
by using the expansion of Rupak and Sch\"afer \cite{rupak} 
we obtain the ETF functional 
(\ref{e-dft}) with $\xi=0.475$ and $\lambda=0.25$. 

Very recently two theoretical groups \cite{chang,blume} have 
studied the two-components Fermi gas 
in the harmonic trap of Eq. (\ref{harmonic}) at unitarity 
by using Monte Carlo algorithms. 
Chang and Bertsch \cite{chang} have used a Green-function 
Monte Carlo (GFMC) method, while Blume, von Stecher and Greene 
\cite{blume} have applied 
a fixed-node diffusion Monte Carlo (FN-DMC) approach. 
They have obtained the ground-state energy for increasing 
values of the total number $N$ of fermions. 

Notice that by inserting the Thomas-Fermi profile (\ref{density-tf}) 
into Eq. (\ref{e-dft}) one gets  the ground-state energy of $N$ fermions 
in a harmonic potential with frequency $\omega$ in the form:
\beq 
{E\over \hbar \omega} = \sqrt{\xi} \Big( {(3N)^{4/3} \over 4} + 
{9\lambda \over \xi } {(3N)^{2/3} \over 8} \Big) \; . 
\label{energy}
\eeq
We will refer to this expression as the "beyond-TF" energy.
The first term is the TF contribution 
to the ground-state energy, while the second term is 
the leading correction.
Chang and Bertsch  \cite{chang} and Blume, von Stecher and Greene 
\cite{blume} have determined $\xi$ by fitting 
their MC data with Eq. (\ref{energy}) 
under the  hypothesis that the relation
\beq 
\lambda = {\xi \over 9} 
\eeq
appropriate to a noninteracting system holds also in the unitary regime. 
They find respectively the values $\xi=0.50$ \cite{chang} and 
$\xi=0.465$ \cite{blume}, which are  compatible with previous 
determinations of the parameter $\xi$ based on Monte Carlo calculations 
for the uniform system\cite{mc1,mc2}. 
The corresponding value $\lambda=\xi/9\simeq 0.05$ for the coefficient of 
the gradient is instead much smaller than previous suggestions 
\cite{kim,manini05,sala-josephson,sala-new,rupak}.

\begin{figure}[tbp]
\begin{center}
\epsfig{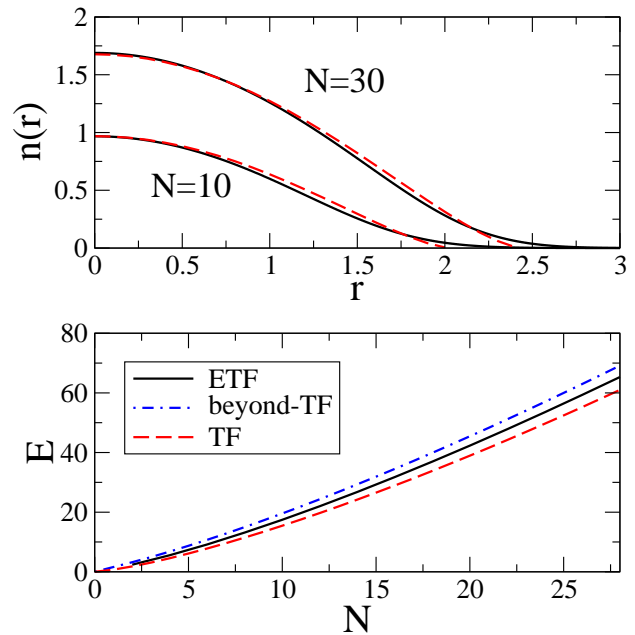}
\end{center}
\caption{(Color online). Unitary Fermi gas 
under harmonic confinement of frequency $\omega$. 
Upper panel: density profiles 
$n(r)$ for $N=10$ and $N=30$ fermions obtained 
with ETF (solid lines) and TF (dashed lines). 
Lower panel: ground-state energy $E$ vs $N$ 
with  ETF (solid line), beyond-TF formula (dot-dashed line) 
and TF (dashed line). In all calculations: 
universal parameter $\xi=0.44$ and gradient coefficient 
$\lambda=1/4$. Energy in units of $\hbar \omega$ and 
lengths in units of $a_H=\sqrt{\hbar/(m\omega)}$.} 
\end{figure}

It is important to stress that the formula in Eq. (\ref{energy}) 
does not give  the minimum of the ETF functional 
(\ref{e-dft}), since it corresponds to the density profile of 
Eq. (\ref{density-tf}) and not to the true ground-state density, 
solution of Eq. (\ref{chem-dft}). 

In the upper panel of Fig. 1 we plot ETF density profiles (solid lines)  
and compare them with the TF ones (dashed lines). 
The ETF density profiles have been determined by solving
Eq. (\ref{chem-dft}) with $\xi=0.44$ 
and $\lambda=1/4$  with a finite-difference 
numerical code \cite{numerics}  As expected, there are visible differences, 
in particular near the surface. 
In the lower panel of Fig. 1 we plot the ground-state 
energy $E$ for increasing values of the number $N$ of fermions. 
Here the differences between TF (dashed line), 
beyond-TF (dot-dashed line) and ETF (solid line) are quite large. 
The figure clearly shows that the beyond-TF formula 
(\ref{energy}) is not very accurate. 

We stress here once again that the values of the parameters $\xi$ and 
$\lambda$ 
in the ETF functional should be universal i.e. independent on the 
confining potential $U({\bf r})$ \cite{ring,parr}. Moreover we consider 
$\lambda$ as taking into account phenomenologically all possible 
corrections of a gradient expansion in the unitary regime and treat 
$\xi$ and $\lambda$ as independent parameters.
To determine them, instead than using the inaccurate Eq.(\ref{energy}), 
we look for the values of the two parameters which lead to the best fit 
of the FN-DMC ground-state energies \cite{blume} for even $N$.
After a systematic analysis we find 
$\xi=0.455$ and $\lambda=0.13$  
as the best parameters in the unitary regime. 
It is important to observe that the value $\xi=0.455$ of our best fit 
coincides with that obtained 
by Perali, Pieri and Strinati \cite{strinati} by using the 
extended BCS theory with beyond-mean-field pairing fluctuations. 

In Fig. 2 we plot the ground-state energy $E$ of the 
Fermi gas under harmonic confinement, comparing 
different results: the FN-DMC data for an even number $N$ of atoms 
of \cite{blume} (diamonds with error bar), 
the best ETF results, with $\lambda=0.13$ and $\xi=0.455$ (solid line), 
and the ETF results obtained by using the values  
$\xi=0.475$ and $\lambda=0.25$ coming from 
the $\varepsilon$-expansion \cite{rupak} (dot-dashed line). 

\begin{figure}[tbp]
\begin{center}
\epsfig{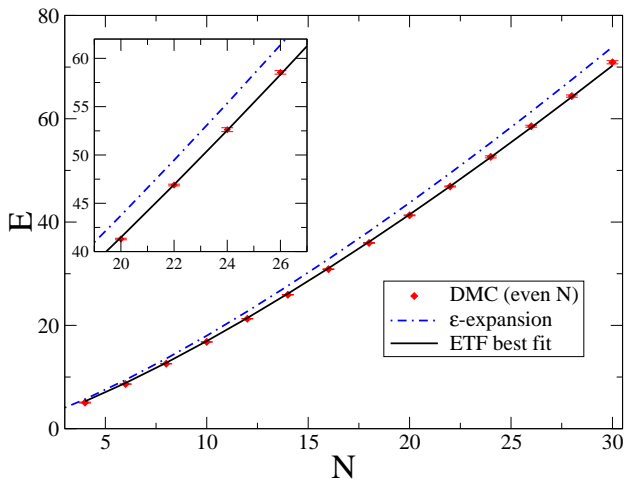}
\end{center}
\caption{(Color online). Ground-state energy $E$ for 
the unitary Fermi gas of $N$ atoms under harmonic 
confinement of frequency $\omega$. Symbols: DMC data with even $N$ 
\cite{blume}; 
solid line: ETF results with best fit ($\xi=0.455$ and $\lambda=0.13$); 
dot-dashed line: ETF results obtained from $\varepsilon$-expansion 
\cite{rupak} ($\xi=0.475$ and $\lambda=0.25$). 
Energy in units of $\hbar \omega$.} 
\end{figure}

We remark that fixing $\xi$ 
to the value $\xi=0.44$ and looking for the best fitting $\lambda$ 
we have found $\lambda=0.18$. In this case the curve of the energy 
will be practically superimposed to the solid one of Fig. 2. 

For the sake of completeness in Table 1 we report the 
fixed node DMC energies of Blume, von Stecher 
and C. H. Greene \cite{blume}, 
our optimized ETF results with $\xi=0.575$ 
and $\lambda=0.13$, and also the SDFT calculations 
of Bulgac \cite{bulgac2}. Remarkably the ETF energies are slightly 
closer to the DMC values than the SDFT ones, reported 
in Ref. \cite{bulgac2}. The optimized ETF energies 
for even number $N$ of particles are obtained from our 
density functional (\ref{e-dft}), while the energies with odd number $N$ 
of particles are calculated taking into account the odd-even splitting,  
as discussed in the next section. 

\begin{center}
\begin{tabular}{|c|c|c|c|}
\hline
~~$N$~~ & ~~$ E_{DMC}$~~ & ~~$ E_{ETF}$~~ & ~~$ E_{SDFT}$~~ \\ 
\hline
  2 &  2.002  & 2.17   &   2.33  \\
  3 &  4.281  & 4.65   &   4.62  \\
  4 &  5.051  & 5.19   &   5.52  \\
  5 &  7.61   & 7.98   &   7.98  \\
  6 &  8.64   & 8.71   &   9.07  \\
  7 &  11.36  & 11.73  &  11.83  \\ 
  8 &  12.58  & 12.61  &  12.94 \\
  9 &  15.69  & 15.81  &  16.06 \\
  10 & 16.80  & 16.83  &  17.15  \\
  11 & 20.11  & 20.19  &  20.36  \\
  12 & 21.28  & 21.32  &  21.63  \\
  13 & 24.79  & 24.82  &  24.96  \\
  14 & 25.92  & 26.04  &   26.32  \\
  15 & 29.59  & 29.67  &  29.78  \\
  16 & 30.88  & 30.99  &   31.21  \\
  17 & 34.64  & 34.73  &  34.81  \\
  18 & 35.96  & 36.13  &   36.27  \\ 
  19 & 39.83  & 39.99  &  40.02  \\
  20 & 41.30  & 41.46  &   41.51  \\
  21 & 45.47  & 45.41  &  45.42  \\
  22 & 46.89  & 46.96  &   46.92  \\
  23 & 51.01  & 51.01  &   $$    \\
  24 & 52.62  & 52.63  &   $$     \\
  25 & 56.85  & 56.76  &   $$    \\ 
  26 & 58.55  & 58.45  &   $$     \\
  27 & 63.24  & 62.66  &   $$    \\
  28 & 64.39  & 64.41  &   $$     \\
  29 & 69.13  & 68.70  &   $$     \\
  30 & 70.93  &  70.51 &   $$     \\
\hline
\end{tabular}
\end{center}
\noindent Table 1. {\small 
Ground-state energy $E$ of the unitary Fermi gas of $N$ even atoms 
under harmonic confinement, in units of the harmonic energy 
$\hbar \omega$. Comparison among different calculations: 
fixed node diffusion Monte Carlo \cite{blume} ($E_{DMC}$), our optimized 
extended Thomas-Fermi with $\xi=0.455$ and $\lambda=0.13$ 
($E_{ETF}$), and the superfluid density functional theory 
\cite{bulgac2} ($E_{SDFT}$). }

\section{Odd-even splitting}

Up to now we have analyzed the unitary gas with 
an even number $N$ of particles. The Monte Carlo calculations 
\cite{chang,blume} show a clear odd-even effect, reminiscent 
of the behavior of the nuclear binding energy. In particular, 
denoting by $E_N$ the ground state energy of $N$ particles 
in a isotropic harmonic trap, for odd $N$ one finds 
\beq 
E_N = {1\over 2} (E_{N-1} + E_{N+1}) + \Delta_N  \; , 
\eeq
where the splitting $\Delta_N$ is always positive. This effect is 
related to pairing: given the superfluid cloud of even particles, 
the extra particle is localized where the energy gap is smallest, 
which is near the edge of the cloud \cite{son2,bulgac2}. 
On this basis, recently Son \cite{son2} has suggested 
that, for fermions at unitarity, confined by a harmonic 
potential with frequency $\omega$, 
the odd-even splitting grows as 
\beq 
\Delta E_N = \gamma \ N^{1/9} \ \hbar \omega \; , 
\label{splitting}
\eeq 
where $\gamma$ is an unknown dimensionless constant. 
After a systematic investigation we find that  $\gamma =0.856$ 
gives the odd-even splitting which best fit entire FN-DMC data, 
with both even and 
odd particles \cite{blume}. 
In Fig. 3 we report 
the FN-DMC data (diamonds) and the 
optimized ETF results (solid line). The figure, which 
 displays the zig-zag behavior of the energy $E$ as a function 
of $N$, shows that the optimized ETF functional plus the 
odd-even correction (\ref{splitting}) ($\xi=0.455$, 
$\lambda=0.13$, $\gamma = 0.856$)  is extremely good 
in reproducing all FN-DMC data.

\begin{figure}[tbp]
\begin{center}
\epsfig{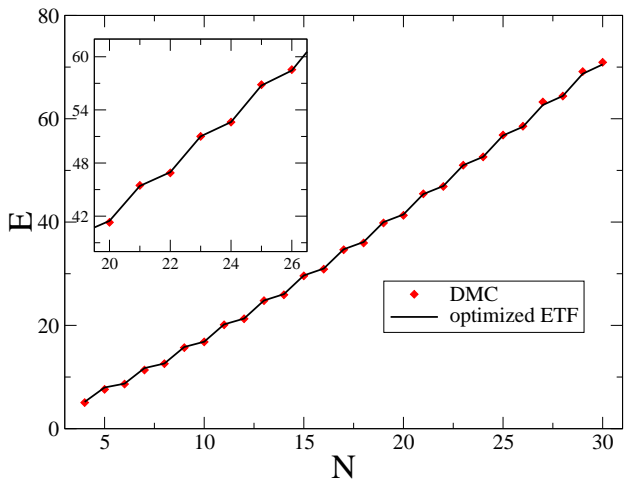}
\end{center}
\caption{(Color online). Ground-state energy $E$ for 
the unitary Fermi gas of $N$ atoms under harmonic 
confinement of frequency $\omega$. Dimonds: DMC data with 
both even and odd $N$ \cite{blume}; solid line: optimized ETF results 
($\xi=0.455$, $\lambda=0.13$, $\gamma=0.856$). 
Energy in units of $\hbar \omega$.} 
\end{figure}

\section{Generalized hydrodynamics}

Let us now analyze the effect of the gradient term (\ref{grad-term})
on the dynamics of the unitary Fermi gas. At zero temperature 
the low-energy collective dynamics of this fermionic gas 
can be described by the equations of generalized 
hydrodynamics \cite{zaremba,tosi,volovik}, where the Hamiltonian of 
the classical hydrodynamics \cite{volovik} is modified by including 
gradient corrections. In our case the generalized Hamiltonian reads:  
\beq  
H = \int d^3{\bf r} \ n \left[ {1\over 2} m v^2  + 
\varepsilon_g(n,\nabla n) 
+ U({\bf r})\right]  \; ,  
\label{ham-g}
\eeq
where the local density $n({\bf r},t)$ and the local 
velocity ${\bf v}({\bf r},t)$ are the hydrodynamics 
variables \cite{tosi,volovik}. By writing the Poisson brackets 
\cite{volovik} of the hydrodynamics variables 
with the Hamiltonian (\ref{ham-g}), one gets the generalized 
hydrodynamics equations:
\beqa 
{\partial n\over \partial t} + \nabla \cdot (n {\bf v}) = 0 \; , 
\label{hy1}
\\ 
m \left( {\partial \over \partial t} + {\bf v}\cdot \nabla\right) {\bf v} 
+ \nabla 
\Big[ \mu_g(n,\nabla n) + U({\bf r}) \Big] = 0 \; .   
\label{hy2} 
\eeqa
where 
\beq 
\mu_g(n,\nabla n) = {\partial 
\big(n \varepsilon_g(n,\nabla n)\big)\over \partial n} =
\mu(n) - \lambda {\hbar^2\over 2m} {\nabla^2 \sqrt{n}\over \sqrt{n}} \; . 
\eeq 
Eq. (\ref{hy1}) is the continuity equation, 
while Eq. ({\ref{hy2}) is the conservation of linear momentum. 
These equations are valid for the inviscid unitary Fermi gas at 
zero temperature. If the unitary Fermi gas is superfluid 
then it is not only inviscid but it is also irrotational, i.e. 
\beq 
\nabla \times {\bf v} = 0 \; .  
\label{irr}
\eeq
By using this condition and the identity 
\beq 
({\bf v} \cdot \nabla) {\bf v} 
= \nabla \left( {1\over 2} v^2 \right) 
- {\bf v} \times ({\bf \nabla} \times {\bf v}) \;, 
\eeq 
Eq. (\ref{hy2}) can be simplified into: 
\beq 
m {\partial {\bf v} \over \partial t} 
+ \nabla 
\left[{1\over 2} m v^2 + \mu_g(n,\nabla n) + U({\bf r}) \right] = 0 \; .   
\label{hy2s}
\eeq
The low-energy collective dynamics of the superfluid Fermi gas 
in the BCS-BEC crossover is usually described by the equations 
of classical superfluid hydrodynamics, which are the time-dependent 
version of the local density approximation with the 
Thomas-Fermi energy functional \cite{stringa-fermi}. 
Equations (\ref{hy1}) and (\ref{hy2s}) are a simple 
generalization of classical superfluid hydrodynamics 
which takes into account surface effects. 
The gradient term, i.e. the 
quantum correction, is necessary in a superfluid to avoid unphysical 
phenomena like the formation of wave front singularities 
in the dynamics of dispersive shock waves \cite{sala-shock}. 

Combining Eqs. (\ref{hy1}) and (\ref{hy2s}) one finds the dispersion 
relation of low-energy collective modes of the 
uniform unitary Fermi gas in the form 
\beq
 {\Omega \over  q} = \sqrt \xi c_0 \sqrt {1+ {\lambda \over \xi} 
\Big( {\hbar q \over { 2 m c_0}}\Big)^2 } \; , 
\label{sound}
\eeq 
where $\Omega$ is the collective frequency, $q$ is the wave number and  
$c_0$ is the speed of sound in a uniform, noninteracting Fermi gas. 

For an irrotational fluid it is possible to write down \cite{volovik,orso}
a Lagrangian by using as dynamical variables the scalar potential 
$\theta({\bf r},t)$ of the velocity ${\bf v}({\bf r},t)$ and 
the local density $n({\bf r},t)$. For a fermionic superfluid one has:
\beq 
{\bf v}({\bf r},t) = {\hbar \over 2m} \nabla \theta({\bf r},t) \; , 
\label{velocity}
\eeq 
where $\theta({\bf r},t)$ is the phase of the condensate wavefunction 
of Cooper pairs \cite{stringa-fermi}. 
In our case, the familiar Lagrangian of the Fermi 
superfluid \cite{son,orso} must be modified by including the gradient 
correction. The generalized Lagrangian density then reads: 
\beq 
{\cal L} = - n \left( {\hbar\over 2} {\dot{\theta}} 
+ {\hbar^2\over 8m} 
(\nabla \theta)^2 + U({\bf r}) + \varepsilon_g(n,\nabla n) 
\right)
\; .  
\label{popov}
\eeq
The Euler-Lagrange equations of this Lagrangian with respect 
of $n$ and $\theta$ give the generalized hydrodynamics equations 
of superfluids (\ref{hy1}) and (\ref{hy2s}). 

We observe that the generalized hydrodynamics equations 
(\ref{hy1}) and (\ref{hy2s}) can be formally written 
in terms of a nonlinear Schr\"odinger equation of the 
Guerra-Pusterla type \cite{guerra}, which is Galilei-invariant 
\cite{doebner}. In fact, by introducing the complex wave function 
\beq 
\Psi({\bf r},t) = \sqrt{n({\bf r},t)}\ e^{i\theta({\bf r},t)} \; , 
\eeq
which is the zero-temperature Ginzburg-Landau order 
parameter normalized to the total number $N$ of superfluid atoms  
\cite{sala-josephson,sala-new}, and taking into account 
the correct phase-velocity relationship given by Eq. (\ref{velocity}), 
the equation 
\beqa 
i \hbar {\partial \over \partial t} \Psi &=& 
\Big[ -{\hbar^2 \over 4 m} \nabla^2 + 2 U({\bf r}) + 
\nonumber 
\\
&&2 2 \mu(|\Psi|^2) +(1 - 4 \lambda ){\hbar^2\over 4m} 
{\nabla^2 |\Psi|\over |\Psi|} \Big] \Psi \; , 
\label{nlse}
\eeqa 
is strictly equivalent to Eqs. (\ref{hy1}) and (\ref{hy2s}). 
Notice that in the stationary case 
where $\Psi({\bf r},t)=\sqrt{n({\bf r})} \ e^{-i2\bar{\mu}t/\hbar}$, 
Eq. (\ref{nlse}) becomes exactly Eq. (\ref{chem-dft}). 
Remarkably only if $\lambda=1/4$ the equation acquires the familiar 
structure of a nonlinear Schr\"odinger such as the Gross-Pitaevskii 
equation  which describes the two-spin component Fermi system, 
but in the extreme BEC regime \cite{dicastro}. 
From the linearization of Eq. (\ref{nlse}) one finds 
for the uniform Fermi gas the Bogoliubov excitations given precisely 
by Eq. (\ref{sound}). 

Finally, we stress that 
Eq. (\ref{nlse}) can be seen as the Euler-Lagrange equation 
of the following Galilei-invariant Lagrangian density 
\beqa 
{\cal L} = \Psi^* 
\left(i {\hbar\over 2} {\partial \over \partial t} + {\hbar^2 \over 8m} 
\nabla^2 - U({\bf r})\right) \Psi 
\nonumber 
\\ 
- \varepsilon_g( |\Psi|^2,\nabla(|\Psi|^2) )|\Psi|^2 + 
{\hbar^2\over 8m} {(\nabla |\Psi|)^2}  \; , 
\eeqa 
that is equivalent to the generalized Lagrangian of Eq. (\ref{popov}). 

\section{Conclusions}

We have obtained the value of the coefficient $\lambda$ of the 
gradient correction $\lambda \hbar^2(\nabla n)^2/(8m n)$
for the extended Thomas-Fermi density functional in the unitary regime. 
By fitting diffusion Monte Carlo data with an even number 
$N$ of particles we have found 
$\lambda=0.13$. In addition, we have determined the coefficient $\xi$
of the energy density $\xi (3/5) n \varepsilon_F$, 
finding $\xi=0.455$. Fixing $\xi$ 
to the value $\xi=0.44$ proposed in \cite{mc2} 
and looking for the best fitting $\lambda$ 
we have found instead $\lambda=0.18$. We stress that 
in our energy functional the gradient term takes
into account phenomenologically all corrections 
of a gradient expansion. Our functional one can 
easily get the ground state properties (energy and density) 
for large as for small numbers of fermions; its 
main limitation is that it cannot account for the shell effects 
in the density profile. Moreover, we have shown that 
it is possible to take into account the odd-even splitting 
of the ground-state energy of the unitary gas in a harmonic trap 
of frequency $\omega$ by considering a correction 
proportional to $N^{1/9} \hbar \omega$ as suggested by Son \cite{son2}, 
where the constant of proportionality is found to be $\gamma=0.856$. 
Finally, we have analyzed the effect of the gradient term in the 
dynamics of the unitary Fermi gas by introducing generalized 
hydrodynamics equations, which can be written for superfluid motion 
in the form of a Galilei-invariant nonlinear Schr\"odinger equation. 
As a final remark, we remind that the values of 
$\xi$ and $\lambda$ we have found are independent 
on the external potential and therefore our generalized energy functional 
can be used to investigate the unitary superfluid Fermi gas in 
various trapping configurations. Obviously in the limit 
of large numbers of particles the gradient term and the 
exact value of $\lambda$ are less important since the dominant 
term becomes the usual Thomas-Fermi one. 
More extensive Monte-Carlo calculations with a larger 
number of particles will be certainly useful 
to have a more accurate determination of the value of $\xi$. 

This work has been supported by Fondazione Cariparo. 
L.S. thanks Sadhan K. Adhikari, Boris A. Malomed and 
Thomas M. Sch\"afer for useful suggestions.


\begin{thebibliography}{99}
\bibitem{nota} It has been recently suggested in Ref. \cite{zyl} 
that for a strictly 2D unitary Fermi gas the gradient corrections 
are not needed: the Thomas-Fermi energy functional should be exact in 2D.
\end{thebibliography}
\end{document}